\documentclass{elsart}
\usepackage{graphics,graphicx,epsfig}
\usepackage{amssymb}

\begin{document}
\begin{frontmatter}

\title{Atomistic mechanisms for the ordered growth of Co nano-dots on Au(788):
comparison of VT-STM experiments and multi-scaled calculations}

\author{S. Rohart},
\author{G. Baudot},
\author{V. Repain},
\author{Y. Girard},
\author{S. Rousset \corauthref{ca}}
\corauth[ca]{Corresponding author} \ead{rousset@gps.jussieu.fr}
\ead[url]{http://ufrphy.lbhp.jussieu.fr/nano/}
\address{Groupe de Physique des Solides: Universit\'{e}s Paris 6 et 7, CNRS, UMR 75-88\\
FR 2437: Mat\'{e}riaux et Ph\'{e}nom\`{e}nes Quantiques, 2 place Jussieu, 75251 Paris Cedex, France}
\author{H. Bulou},
\author{C. Goyhenex}
\address{Institut de Physique et Chimie des Mat\'{e}riaux de Strasbourg, UMR 75-04, 23 rue du Loess, 67037
Strasbourg, France}
\author{L. Proville}
\address{Service de Recherches de M\'{e}tallurgie Physique CEA/Saclay, 91191 Gif sur Yvette Cedex, France}

\begin{abstract}
Hetero-epitaxial growth on a strain-relief vicinal patterned
substrate has revealed unprecedented 2D long range ordered growth
of uniform cobalt nanostructures. The morphology of a Co
sub-monolayer deposit on a Au(111) reconstructed vicinal surface
is analyzed by Variable Temperature Scanning Tunneling Microscopy
(VT-STM) experiments. A rectangular array of nano-dots (3.8~nm x
7.2~nm) is found for a particularly large deposit temperature
range lying from 60~K to 300~K. Although the nanodot lattice is
stable at room temperature, this paper focus on the early stage of
ordered nucleation and growth at temperatures between 35 K and 480
K. The atomistic mechanisms leading to the nanodots array are
elucidated by comparing statistical analysis of VT-STM images with
multi-scaled numerical calculations combining both Molecular
Dynamics for the quantitative determination of the activation
energies for the atomic motion and the Kinetic Monte Carlo method
for the simulations of the mesoscopic time and scale evolution of
the Co submonolayer.
\end{abstract}

\end{frontmatter}

\section{Introduction}

Nucleation and growth of mono-disperse nanostructures is a
challenging field both for theoretical modeling and practical
applications due to their new magnetic, electric and catalytic
properties. Growth of regular islands has been achieved in various
systems such as metal aggregates supported on insulator surfaces
\cite{haas2000,heim1996}, hetero-epitaxial growth of
semiconductors including self-assembled quantum dots
\cite{tersoff1996,teichert2002}, and metal on metals systems
\cite{chambliss1991,brune1998b}. Although nucleation and growth
models have been extensively developed and compared with
experiments in the case of homogeneous substrates, growth on
heterogeneous substrate has been considered only recently. The
hetero-epitaxial growth of highly strain islands is still very
difficult to modelize, but simple systems have been considered
such as nucleation on substrates containing point-defects traps
\cite{haas2000,heim1996,venables2003} or spatial ordering of
islands grown on patterned substrates \cite{brune1998b,lee1998}.
The use of spontaneously nanostructured substrates as templates
for organized growth is a promising way as it allows to grow not
only regular nanostructures but also to grow high density lattice
of regular nanostructures over macroscopic scales. This opens up
the way for new studies of both individual and collective physical
properties by means of standard averaging technics.

Metal on metal systems provide model systems for ordered growth on
well-defined nano-patterned substrates
\cite{chambliss1991,brune1998b,ellmer2002}. Experimentally, this
phenomena has been successfully applied to the formation of
nanostructures but very few studies are dealing with the atomistic
mechanisms leading to the organization and the quantitative
determination of the associated energies. The precise
determination of the atomistic mechanisms for a given substrate
should allow to make some prediction to get an ordered growth with
various deposited material and to find out the conditions (flux,
temperature) for the narrowest size distribution. Indeed the use
of a nanostructured template is not a sufficient condition to get
an ordered growth. For example, the reason why Fe, Co, Ni, Cu
\cite{chambliss1991,voigtlander1991a,voigtlander1991b,bulou2001}
display an ordered growth on Au(111) at room temperature whereas
Ag, Al \cite{bulou2001,meyer1996,fischer1999} do not is still
under debate.

In this paper we study the nucleation and growth of Co nanodots on
a Au(111) reconstructed, vicinal substrate. This system has been
shown recently \cite{repain2002a,repain2002b,baudot2003} to
display an improved long-range order and a narrower size
distribution than on Au(111). Using a Variable Temperature
Scanning Tunneling Microscope (VT-STM), we have studied the
evolution of the growth morphology with temperature. With a
statistical analysis of the STM images combined with Kinetic Monte
Carlo (KMC) simulation and Molecular Dynamics (MD) calculation, we
determine the mechanisms at the origin of the organization of Co
nanodots on Au(788). This result opens the way to make
quantitative predictions about the ordered growth of other
materials on Au(788) like Fe or Ni.

\section{Experimental Details}

The Au(788) substrate is a stable, vicinal surface misoriented by
$3.5^{\circ}$ with respect to the (111) plane toward the [-211]
azimuth. It has been fully presented in a previous paper
\cite{repain2002a}. Its surface displays a very regular succession
of mono-atomic steps and $3.8$ nm wide terraces. Due to the
$22\times \sqrt{3}$ reconstruction of the Au(111) plane, Au(788)
is also structured in the direction perpendicular to the steps
(7.2 nm periodicity). It was shown in earlier studies that this
surface can be used as a template for the growth of cobalt
nano-dots \cite{repain2000,repain2002a,repain2002b,baudot2003} as
the crossing of a discommensuration line and a step edge acts as a
favored nucleation site.

Our Au(788) sample is a Au single crystal cut by spark erosion to
produce a 4 mm diameter and 2 mm thick disk. It was prepared in a
UHV chamber (base pressure $3 \times 10^{-11}$ mbar) using
repeated argon sputtering ($5 \times 10^{-6}$ mbar Ar pressure,
$1$ keV energy) followed by 850 K annealing cycles.

The cobalt is evaporated from a $2$ mm diameter cobalt rod
directly heated by electron bombardment ($I_{em} = 12$ mA, $HV =1$
kV). The pressure during this process is always below $2\times
10^{-10}$ mbar. The flux rate is about $0.2$ ML per minute. The
uncertainty on the absolute cobalt coverage is about $20 \%$.
Deposition was achieved directly under a VT-STM. For temperatures
below room temperature, STM images were taken at the growth
temperature in order to be sure that no change could be introduced
by the annealing. For deposition temperatures above room
temperature, STM images were taken at room temperature as cooling
down the sample does not change the result (clusters density, size
distribution and clusters morphology).

\section{VT-STM experiments}

In order to determine the atomistic mechanisms at the origin of
the ordered growth of Co on Au(788), we have performed STM images
for different substrate temperatures and Co coverage. In a first
part, we present and describe the STM images for temperature
ranging from 35 K to 430 K. In a second part, we extract mean
quantities from a statistical analysis of the images.

\subsection{STM images}

\begin{figure}[ht]
  \begin{center}
    \includegraphics[width=13cm]{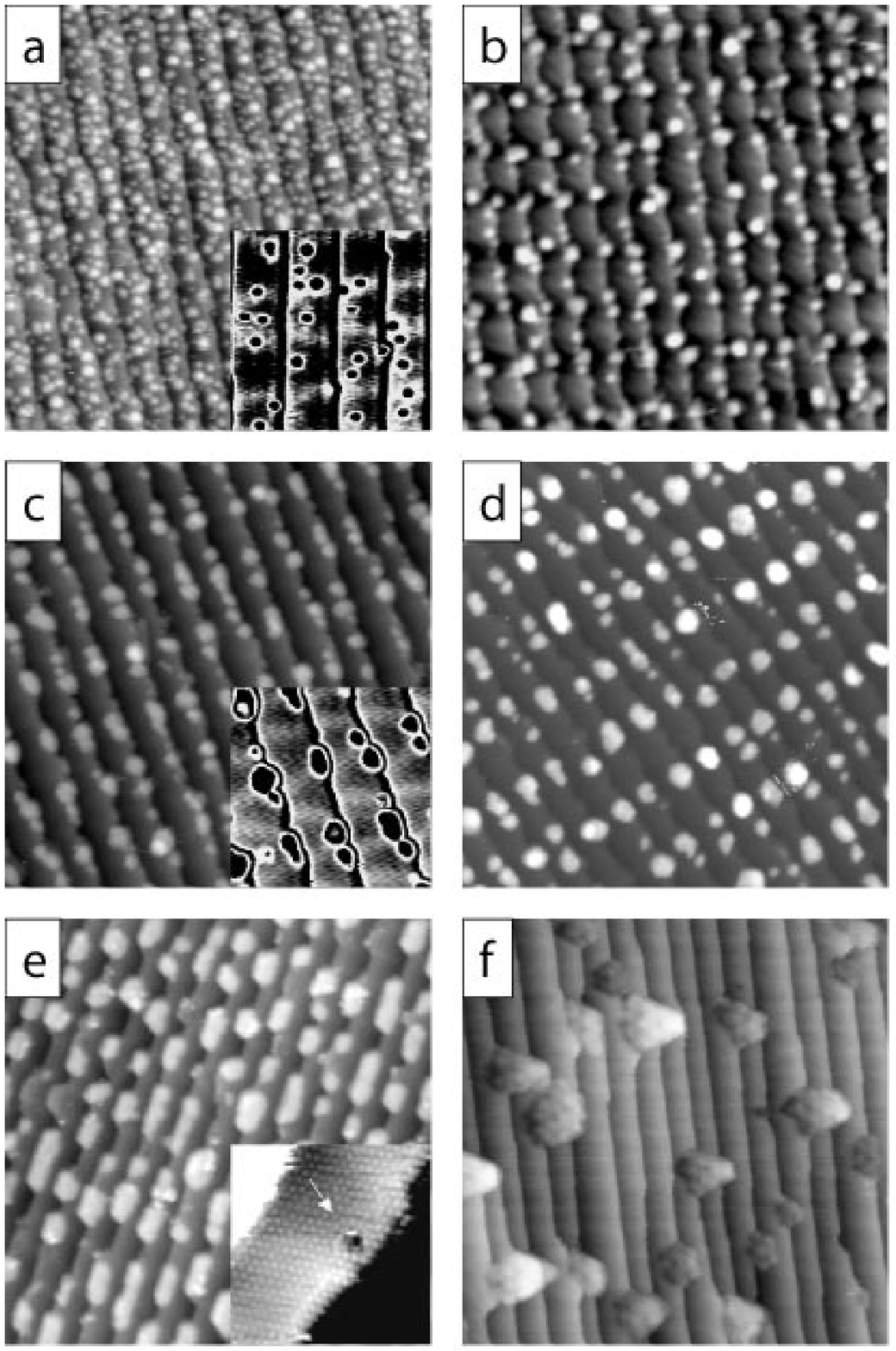}
    \caption{STM images of cobalt deposition on Au(788) for
      different temperatures. Every image is $50$ nm wide.
      a) T = 40 K, $\theta=0.6$ ML: the dots are randomly distributed on the
      surface. (inset: 16 nm wide image of 0.005 ML Co on Au(788). The contrast due to
      the steps was substracted to enhance the reconstruction (white
      lines).)
      b) 65 K, $\theta=0.3$ ML: a rough organization is obtained.
      c) 95 K, $\theta=0.3$ ML: a good organization is obtained.
      Dots are located at the crossing of a discommensuration line
      and a step edge. (inset: 16 nm wide detail of the STM image.
      The contrast due to the steps was substracted in order to enhance
      the reconstruction.)
      d) 170 K, $\theta=0.4$ ML: the good organization in c) is
      maintained for temperature up to 170 K.
      e) 300 K, $\theta=0.3$ ML: the organization on the surface
      disappears and lots of inhomogeneities are seen.  (inset:
      atomically resolved 8 nm wide image which shows an inserted
      nuclei of Co near the crossing of the discommensuration line
      and the step edge. This phenomena was observed for temperatures
      above 200 K.)
      f) 430 K, $\theta=0.4$ ML: the dots are randomly distributed on the surface.}
      \label{STMimg}
  \end{center}
\end{figure}

STM images of the Co deposition on Au(788) at different
temperatures are presented in figure \ref{STMimg}. Roughly, three
different regimes are found: first, for low temperatures (below 60
K, see fig. \ref{STMimg}a) no order is found: small clusters are
randomly dispersed on the surface and the clusters density is very
high. The inset of fig. \ref{STMimg}a shows a low Co coverage
image, where single Co adatoms can be seen. We can notice that the
positions of these adatoms seem not to be influenced by the
surface structuration (steps, discommensuration lines). At 65 K
(fig. \ref{STMimg}b), an organization clearly appears. However,
many islands are still randomly located on the surface. Above 65
K, the quality of the organization improves a lot. Few evolution
is seen versus the temperature: the same result is obtained from
95 K to 170 K. In the inset of \ref{STMimg}c, we can see that dots
are located near the crossing of a discommensuration line and a
step edge. We call these sites "favored nucleation sites". As a
consequence an array of pairs of dots placed on a rectangular
lattice is formed on the surface. This array displays a long range
order as there is exactly one dot on every favored site. Only few
defects (missing dot, dot in a non preferential surface site...)
can be seen, which shows the quality of the array. From 170 K to
300 K the order remains but it is not as good as in the previous
temperature range and an increasing number of defects (missing
dots, coalesced neighboring dots) can be seen. From 200 K to 300
K, low Co coverage images show that many Co atoms are inserted in
the gold surface layer in preferential atomic sites located near
the step edges and the discommensuration lines (cf. inset of fig
\ref{STMimg}e). For temperatures higher than room temperature, it
has been shown \cite{repain2002a,repain2002b} that the order
disappears. We can see on fig. \ref{STMimg}f that the growth
result is a few, big, faceted clusters randomly distributed on the
surface.

\begin{figure}[ht]
  \begin{center}
    \includegraphics[width=16cm]{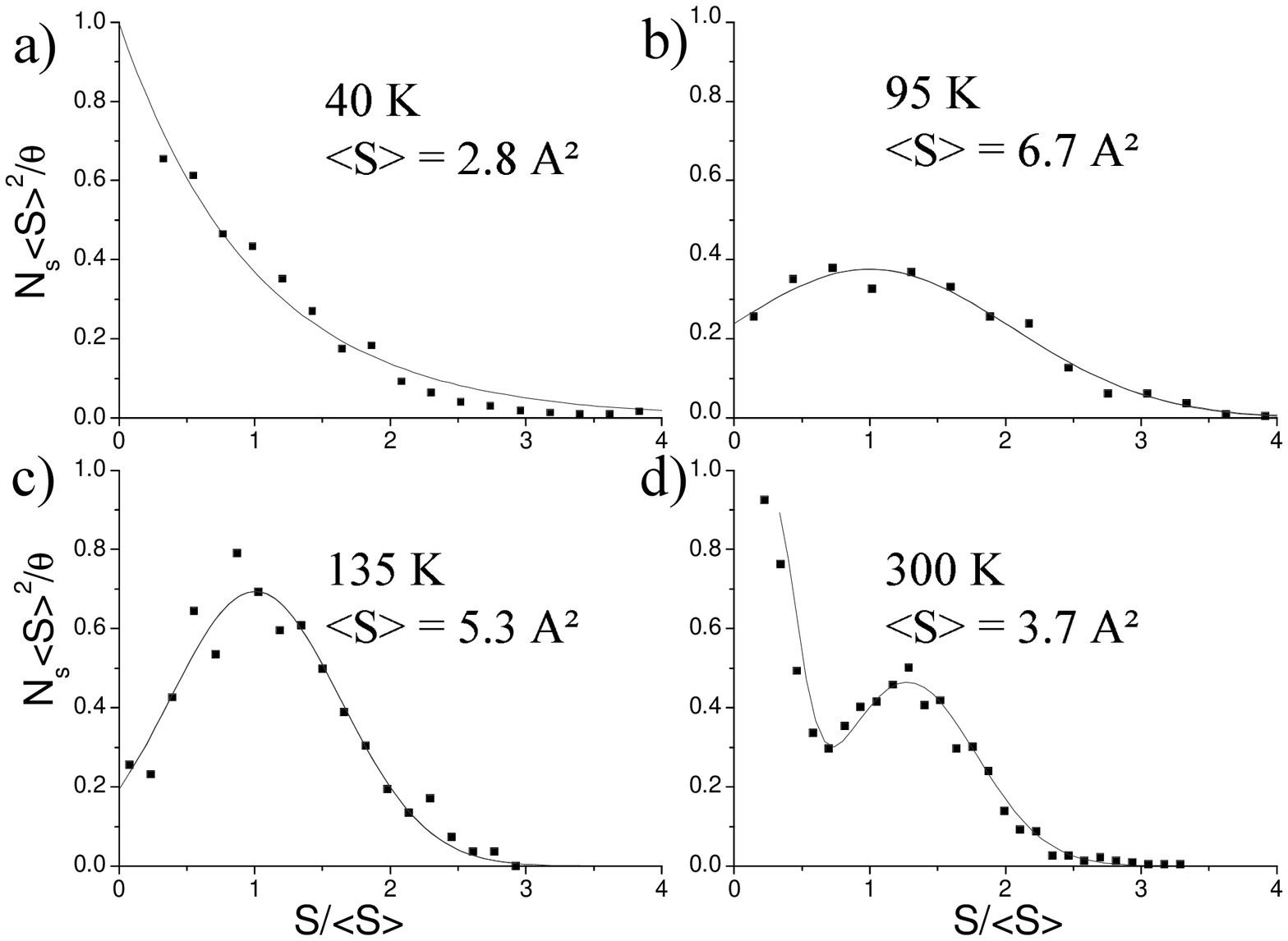}
    \caption{\label{size_dis} Normalized size distribution of the Co
      clusters for different
      temperatures. a) T = 40 K, $\theta = 0.6$ ML, fitted by an
      exponential decay, b) T = 95 K, $\theta=0.3$, fitted by a
      gaussian, c) T = 135 K, $\theta=0.3$, fitted by a gaussian and d)
      T = 300 K, $\theta=0.3$, fitted by a combination of a gaussian and an exponential
      decay.}
  \end{center}
\end{figure}

The size distributions for different substrate temperatures are
presented in figure \ref{size_dis}. These size distributions were
performed with an object processing software on several images
(five to ten 60 nm wide STM images i.e. 1000 to 2000 dots). At low
temperature (40 K in the figure \ref{size_dis}a), the size
distribution does not show a maximum and the smallest clusters are
the most often found. For higher temperatures (95 K and 135 K,
figures \ref{size_dis}b and c), the size distributions show a
maximum. Moreover, in the 95 K - 170 K range, the width at half
maximum decreases when temperature increases and rather good
mono-disperse size distributions are obtained above 135 K. It is
worth to notice that an annealing to room temperature increases
dramatically the quality of this size distribution
\cite{repain2002a,baudot2003}. At 300 K, the size distribution
shows that two kinds of dots co-exist on the surface: some small
ones (less than 20 atoms) and some greater ones (more than 100
atoms). This is in agreement with the STM image shown in figure
\ref{STMimg}e which reveals lots of inhomogeneities. The origin of
these inhomogeneities will be discussed later with the help of the
KMC simulations. \\

\subsection{Critical density study}

\begin{figure}[ht]
  \begin{center}
    \includegraphics[width=16cm]{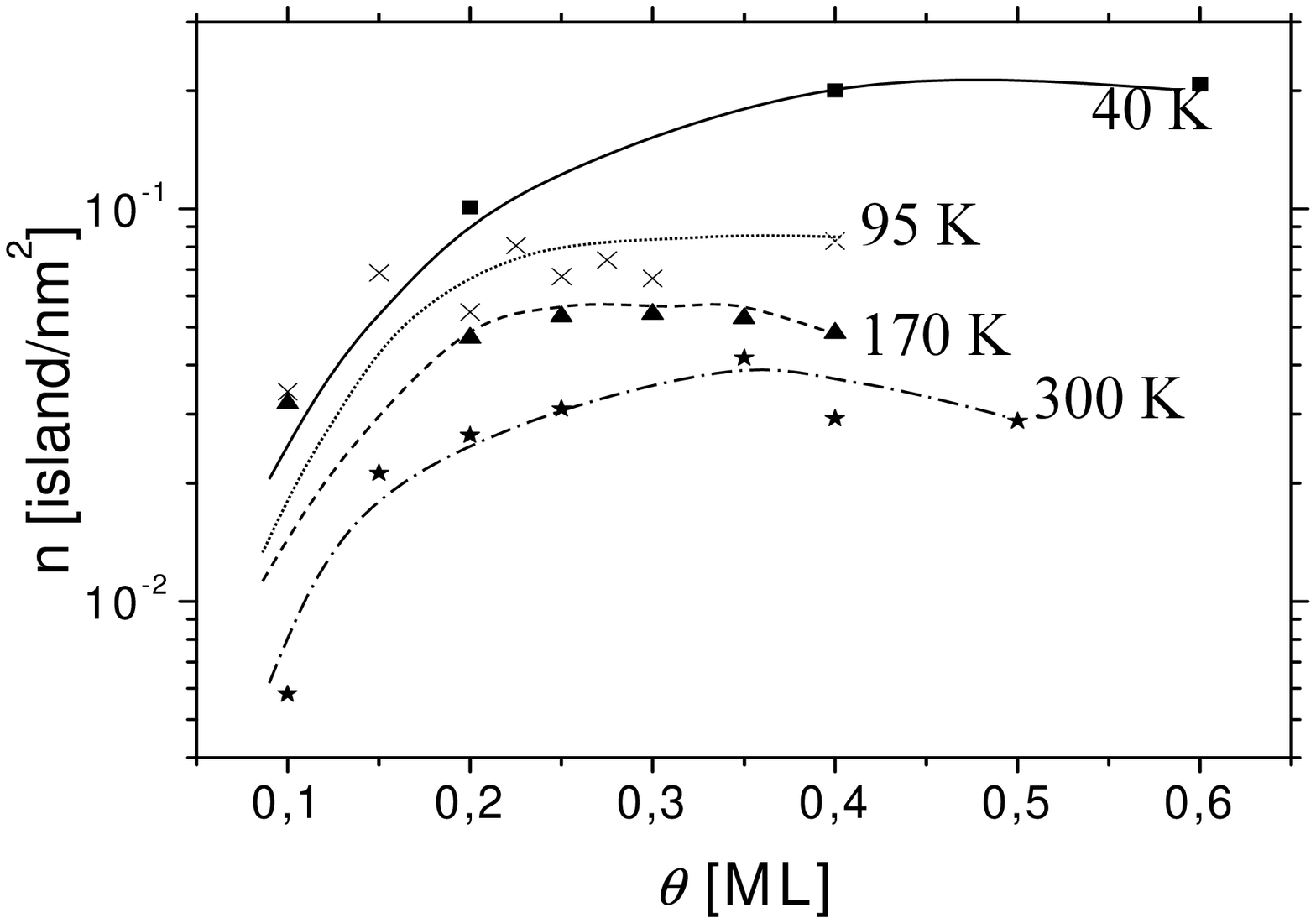}
    \caption{\label{n_vs_th} The clusters density is plotted versus the Co
      coverage ($\theta$) for different substrate temperatures.
      The dots are the experimental data and the lines are guides for
      the eyes.}
  \end{center}
\end{figure}

For every substrate temperature (40 K - 480 K), cobalt deposition
was done step by step in order to record the evolution of the
growth morphology with the coverage. We have extracted from these
images the Co clusters density as a function of the coverage. Some
of these curves for several temperatures are presented in figure
\ref{n_vs_th}. This evolution reflects the well-known scenario of
the sub-monolayer growth on surfaces (see for example
\cite{brune1999} or\cite{jensen1994}). At the beginning the
density of clusters increases regularly as the adatoms nucleate
new clusters. Then the density of clusters stabilizes and the size
of the clusters increases: in this regime the density is constant
and is called the critical density ($n_c$). For higher coverage,
the coalescence regime is reached and the density decreases.

Several studies
\cite{brune1999,venables1972,venables1987,bott1996,brune1998a}
have already pointed out the importance of the critical clusters
density (i.e. the maximum cluster density versus the coverage.),
which reflects the diffusion length of adatoms on the surface, to
elucidate the growth mechanisms.

\begin{figure}[ht]
  \begin{center}
  \includegraphics[width=16cm]{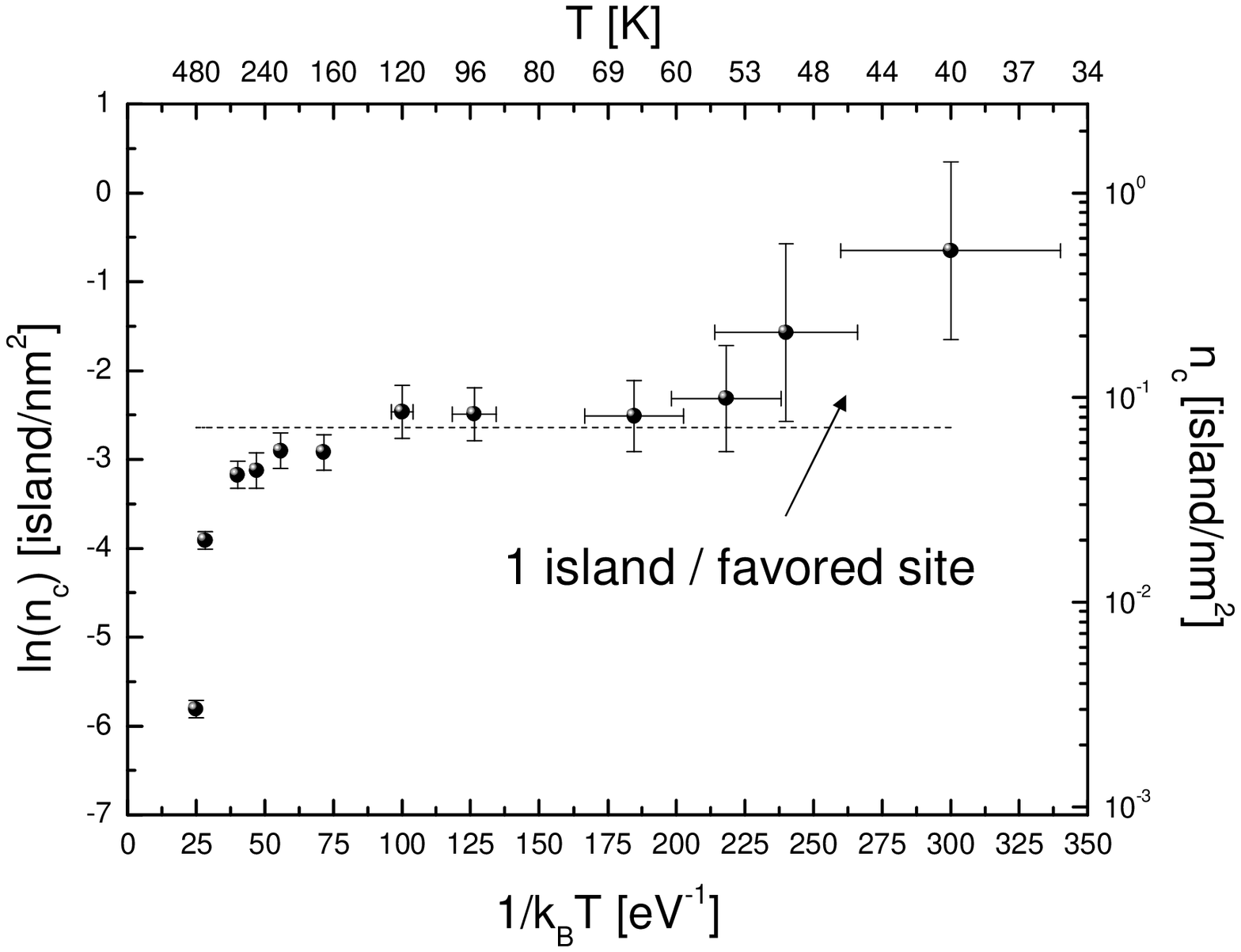}
  \caption{ \label{n_vs_T} Arrhenius plot of the critical clusters density
      evolution with temperature. The dotted line indicate the density
      of favored nucleation sites on Au(788).}
  \end{center}
\end{figure}

The evolution of the critical cluster density versus the
temperature is given in figure \ref{n_vs_T} in a standard
Arrhenius plot. This curve gives a quantitative analysis of the
STM images presented before. It displays a large range of
temperature (60 K - 300 K) where the density of clusters is nearly
constant. This clearly shows that the growth is not homogeneous on
the Au(788) surface which is explained by the nano-patterning of
the surface. We can remark that the density equals the density of
favored nucleation sites, which is in agreement with STM images
described above (fig. \ref{STMimg}c and d). Once more the curve
clearly shows that the quality of the dot array decreases above
160 K as the clusters density slowly decreases until 300 K as
pointed out above. Under 60 K and above 300 K, the strong linear
decrease of the logarithm of the clusters density as a function of
$1/k_BT$ is typical of nucleation and growth on homogeneous
surfaces \cite{brune1998a}. This indicates that the growth is less
influenced by the favored nucleation sites.

\section{Interpretation and multi-scaled calculations}

We have demonstrated that Co displays an ordered growth on Au(788)
in a large temperature range (60~K - 300~K). As a comparison, one
can see that it is particularly larger than the plateau found by
H. Brune et al. \cite{brune1998b} for the organized growth of Ag
nanodots on Ag bilayer on Pt(111) ($100$ K - $130$ K). In this
case the organized growth was due to a confinement of the adatoms
by the dislocation lines of the strain-relief pattern. The
organized regime stops as soon as the adatoms have sufficient
energy to cross these lines. We wonder which atomistic mechanisms
are at the origin of the ordered growth on Au(788) and explain
such a large temperature range.

\subsection{A simple view of the organized growth}

What are the pertinent parameters, which explains the temperature
range for an ordered growth regime ?

For low temperatures, the mean free path of adatoms on a surface
is lower than the mean distance between favored sites (or the
density of cluster is higher than the density of favored sites as
it can be seen on fig. \ref{n_vs_T}). In consequence, the adatoms
cannot be influenced by the periodic patterning of the surface as
its period is wider than the region explored by adatoms. That is
why no order is found for a growth with such substrate
temperatures. We can estimate the temperature threshold ($T_{o}$)
overwhich the system displays an ordered growth as the temperature
at which the diffusion length of the adatoms on an hypothetical
homogeneous substrate equals the mean distance between two favored
sites. As a consequence, the parameters that determine the
beginning of the ordered growth regime are the diffusion energy
$E_{diff}$ and the mean distance between favored nucleation sites
$l_{t}$ (or, what is more convenient, the density of favored sites
$n_{t}=1/l_{t}^2$).

Using the well-known Rate Equation (RE) model for homogeneous
growth \cite{venables1972,bott1996,brune1998a,villain1992}, we
estimate the lowest temperature $T_o$, for which the ordered
growth regime is reached for Co on Au(788). We assume that all
dimers are stable on the surface (critical cluster size $i^*=1$),
as for low temperature the dimer cohesion energy is much higher
than the thermal energy. The evolution of the critical clusters
density with temperature is given by:

\begin{eqnarray}\label{i=1}
  n_c & = & \eta (D/\sigma F)^{-1/3} \nonumber\\
      & = & \eta (D_0/\sigma F)^{-1/3}\exp(E_{diff}/3k_B T)
\end{eqnarray}

with $\eta$ a fixed prefactor (about 0.25 using the lattice
approximation for capture rates \cite{brune1998a}), $F$ the
deposition rate (Flux), $D_0$ the diffusion coefficient at 0 K,
$\sigma$ the size of a lattice site and $E_{diff}$ the diffusion
energy. The ordered growth regime is reached when the clusters
density equals the density of favored sites ($n_t$). Taking
$n_c=n_t=1/200$ the ratio of favored sites per atomic sites,
$E_{diff} = 0.12$ eV \cite{goyhenex2001,liu2002} and
$D_0=5.8\times10^{12}$ \AA$^2$/s \cite{bulou2003} and $\sigma=7.2$
\AA$^2$ we find that an ordered growth should be observed for
temperatures over 83 K, which is a little higher than the
experimental value ($T_o^{STM}=70$ K). As we know that the RE
model given in equation eq. \ref{i=1} tend to over estimate the
dot density \cite{bott1996}, we find that there is a good
agreement between this qualitative estimation and STM experiments.

For temperatures above $T_{o}$, some atoms are trapped into the
favored sites and we observe an ordered growth regime. A
characteristic of such a regime is that the critical clusters
density is constant with temperature and equals the density of
favored sites. Indeed the diffusion length of the adatoms is
limited by the presence of the favored sites. This ordered growth
occurs as long as the typical energy of the trapping mechanism is
sufficient to stabilize adatoms in the favored sites. We call
$T_{e}$ the last temperature for which an ordered growth is
observed. The most important parameter, which determine $T_{e}$ is
the energy barrier for an atom to jump out of the trap $E_{t}$.

In order to estimate $T_{e}$, we modify the RE model for
homogeneous growth to take into account the favored sites,
following the idea of J. Venables in
\cite{venables2003,venables1997}. Assuming for simplicity that
dimers are stable on the surface (the dimers binding energy is
infinite), the RE for the homogeneous growth describe the
evolution of the population of adatoms $n_1$ and stable island
$n_x$. In this new model, we add an homogeneous distribution of
traps with the density $n_t$. We then need to consider the
evolution of the population of trapped adatoms $n_{1t}$ and
trapped stable islands $n_{xt}$. The evolution of $n_{1t}$ is
mainly given by the rate for an adatom to reach a trap $\sigma_1 D
n_1 n_{te}$ and by the rate for this atom to jump out of the trap
$n_{1t}\nu_0\exp(-E_t/k_BT)$ \cite{venables1997}. $\sigma_1$ is
the capture rate of trap (which is assumed for simplicity to be
equal to the capture rate of an adatom), $D$ is the diffusion
coefficient ($D=D_0\exp(-E_{diff}/k_BT)$), $n_{te}$ is the density
of empty traps ($n_{te}=n_t-n_{1t}-n_{xt}$), $\nu_0$ is the
attempt rate and $E_t$ is the energy barrier for the jump. The
four rate equations for the evolution of $n_1$, $n_{1t}$, $n_x$
and $n_{xt}$ are given by :

\begin{equation}\label{rate}
  \left\{
\begin{array}{rcl}
  \frac{dn_{1}}{dt}  & = & F - \sigma_1 D n_1(2n_1+n_{te}+n_{1t})-\sigma_x D n_1(n_x+n_{xt})+n_{1t}\nu_0\exp(-E_t/k_BT)\\
  \frac{dn_{1t}}{dt} & = & \sigma_1 D n_1(n_{te}-n_{1t})-n_{1t}\nu_0\exp(-E_t/k_BT)\\
  \frac{dn_{x}}{dt}  & = & \sigma_1 D n_1^2\\
  \frac{dn_{xt}}{dt} & = & \sigma_1 D n_1 n_{1t}
\end{array}
  \right.
\end{equation}

\begin{figure}[ht]
  \begin{center}
    \includegraphics[width=16cm]{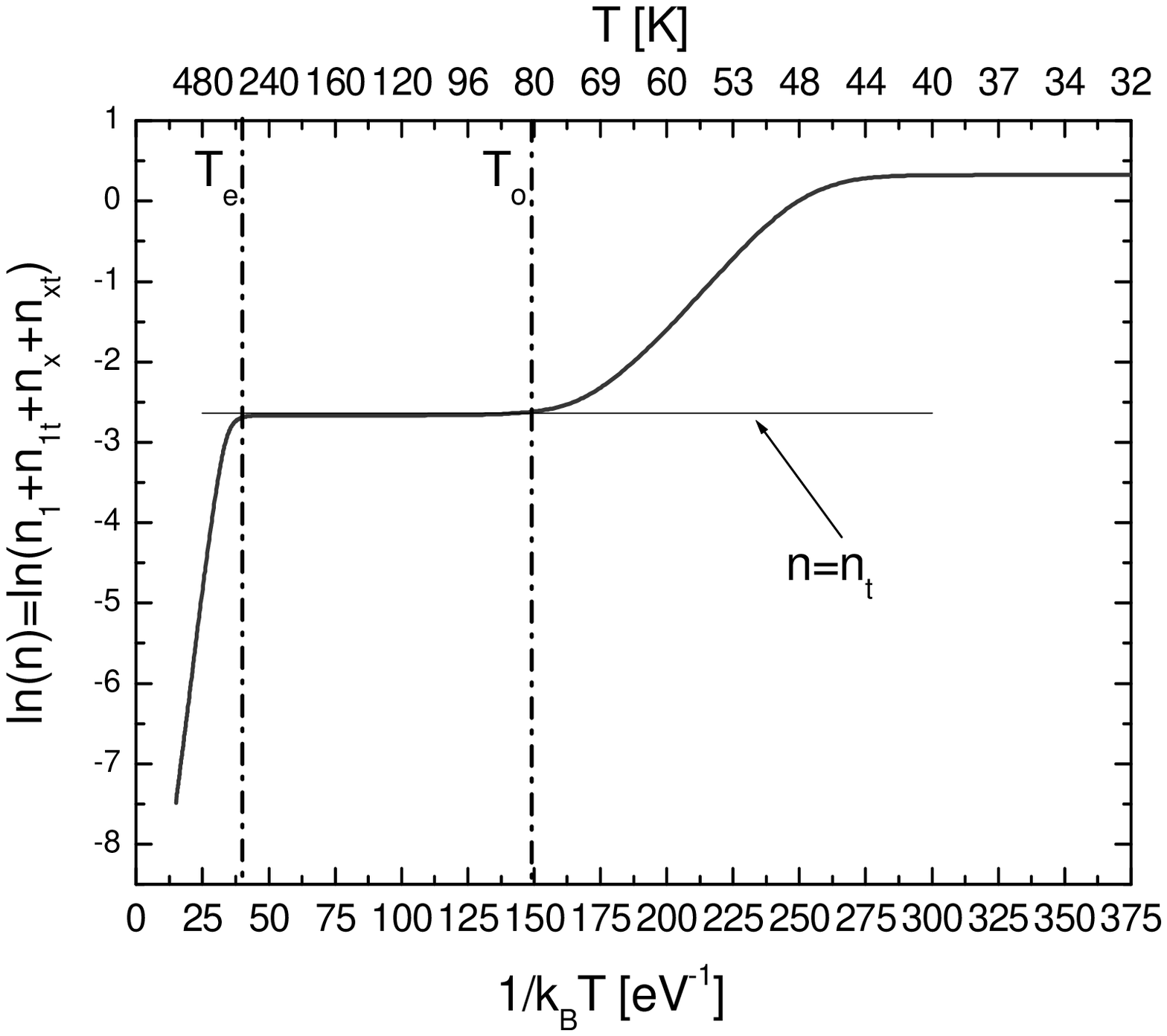}
    \caption{\label{figMF} Rate Equation calculation. The model take into
      account traps on the surface. The energy to leave the traps is set
      to 0.82 eV in order to fit the experimental data.}
  \end{center}
\end{figure}

A key parameter of these equations is the capture rates for an
adatom ($\sigma_1$) and a stable cluster ($\sigma_x$). The choice
for the values is discussed by H. Brune in \cite{brune1998a}. We
take $\sigma_1=3$ and $\sigma_x=7$, which was found to give a good
agreement for low coverages. In these equations, we neglect the
possibility of direct impingement of a deposited atom onto
monomeres or islands. The numerical integration of these equations
allows us to determine the evolution of the cluster density on the
surface. Taking $E_t$ as a free parameter, the figure
fig.\ref{figMF} gives the best fit of the experimental data for
the evolution of the cluster density with the temperature. With
$E_t=0.82$ eV, the model in eq.\ref{rate} reproduce very well the
position of the end of the plateau with $T_e\simeq300$ K. This
result is consistent with previous theoretical
\cite{venables2003,venables1997} and experimental studies of the
growth of $Pd$ on $MgO$ \cite{haas2000} and $Fe$ on $CaF_2$
\cite{heim1996} where point defects on the surface trap the
adatoms on the surface with leads to a favored nucleation on the
point defects.

These RE calculation gives an estimation of the key parameters
involved in the growth of Co on Au(788). However, in order to go
further than the mean-field approximation and to take into account
more complex mechanisms we perform KMC simulations together with
MD calculations for the quantitative determination of the
activation energies of the mechanisms involved in the KMC.

\subsection{Multi-scaled calculations}

\subsubsection{Principle of the calculations}

The KMC modeling is a powerful tool for studying the time
evolution of the growth of atom clusters over the time scale of
diffusion and for mesoscopic space scale. About theory of the KMC
for the growth of thin film, the reader can refer to the works of
P. Jensen in \cite{jensen1994} and H. Brune in \cite{brune1998a}.
The comparison between the STM images and the simulations gives an
accurate insight about the atomic mechanisms that yield the
ordered growth of Co nanodots.

We have developed a KMC code based on the algorithm of Bortz,
Kalos and Lebowitz \cite{bortz1975}. The surface is modeled by an
array of adsorption sites placed on an hexagonal lattice. Each of
this site represents an atomic position where a Co atom may reach
a local mechanical equilibrium. The atomic displacements between
those sites are ruled by the transition state theory:

\begin{equation}\label{pi}
  p_i=\nu_0 \exp(-\Delta E_i/k_B T)
\end{equation}

The attempt frequency $\nu_0$  is fitted so that the KMC model
reproduce the correct diffusion coefficient $D_0$ given in
\cite{bulou2003}. We then have $\nu_0=D_0/(4\sigma)$ with $\sigma$
the surface of an atomic site ($\sigma = 2.88\times2.88\sqrt{3}/2
= 7.2$ \AA$^2$).

$\Delta E_i$ is the energy barrier for the atomic  processes. In
principle the number of different barriers is very large, and
their exact values are unknown. Only approximate total energy
methods can at present give results for all the processes of
interest here. We have used the Quenched-Molecular-Dynamics (QMD)
to give us an idea about the important processes. QMD is an energy
minimization procedure, based on the integration of the equation
of motion for each atom in the system, which consists of canceling
the velocity of the atoms when the product of the force acting on
the atoms by their velocity becomes negative. Then the kinetic
energy of the system decreases leading to the minimization of the
potential energy at 0 K \cite{luo1988,bennett1975}. In our
calculations, the equations of motion are integrated with a
velocity-Verlet integrator \cite{swope1982} with a time step of 5
fs which is the best compromise between calculation speed and
system stability. We considered that the quenched situation is
reached when the system temperature is lower than 3 mK whith
temperature fluctuations lower than 0.2 $\mu$K/fs, which ensure an
energy precision better than 0.1 meV.

The interatomic forces are calculated in the framework of the
Second Moment Approximation of the Tight Binding Theory (TBSMA)
\cite{ducastelle1970} from the total energy
\begin{equation}
  \label{eq:Etot}
  E_{tot}=\sum_i\left\{\sum_{j\neq i}A_{X_iY_j}\exp\left[-p_{X_iY_j}\left[\frac{r_{ij}}{r^{X_iY_j}_0}-1\right]\right]-\sqrt{\sum_{j\neq i}\xi_{X_iY_j}^2\exp\left[-2q_{X_iY_j}\left[\frac{r_{ij}}{r^{X_iY_j}_0}-1\right]\right]}\right\}
\end{equation}
$X$ and $Y$ indicate the chemical species (Co,Au), $r_0^{XX}$ the
first-neighbour distance in the metal $X$ and
$r_0^{XY}=\frac{1}{2}(r_0^{XX}+r_0^{YY})$. The parameters
$A_{XY}$, $q_{XY}$, $p_{XY}$, and $\xi_{XY}$ (Table
\ref{tab:parametre}) are fitted to the experimental values of the
cohesive energy, lattice parameter, and elastic constants
\cite{simmons1971,kittel1996} for homoatomic interactions
($Co-Co$, $Au-Au$). TBSMA potentials are known to underestimate
surface energies. Then,  a peculiar attention has been paid in
order to reproduce the difference between the surface energies of
Co and Au \cite{goyhenex2001}. Heteroatomic interaction $Au-Co$
parameters are calculated by fitting the positive heats of
solution for a single substitutional impurities and refined in
order to reproduce the existence of a miscibility gap in the phase
disgram of the bulk $Au-Co$ system \cite{chado2003}.

\begin{table}[htbp]
  \begin{center}
    \begin{tabular}{l|ccccc}
      \hline      \hline
      Element & $A$(eV) & $p$   & $\xi$(eV) & $q$   & $r_0$(\AA) \\
      \hline
      \hline
      Au-Au   & 0.189   & 10.40 & 1.744     & 3.87  & 2.880    \\
      Au-Co   & 0.140   & 10.63 & 1.656     & 3.11  & 2.695     \\
      Co-Co   & 0.106   & 10.87 & 1.597     & 2.36  & 2.510    \\
      \hline      \hline
    \end{tabular}
    \caption{Parameter value for $Au-Au$, $Au-Co$, and $Co-Co$ interactions}
    \label{tab:parametre}
  \end{center}
\end{table}

Calculations have been performed on an Au(111) slab consisting of
two terraces, each of them containing  $(10\times 15\times 8)$
gold atoms, plus one cobalt adatom, with periodic boundary
conditions in the directions $<1\bar{1}0>$ and $<\bar{1}\bar{1}2>$
parallel to the surface. The dimension of the terraces along the
$<\bar{1}\bar{1}2>$ has been set in order to reproduce the
dimensions of the Au(788) terraces.

The determination of the relevant processes needs to calculate the
activation energy of the different possible mechanisms, i.e. by
calculating the energy of the system along the minimum energy path
between initial and final states. The problem of finding the
minimum energy path for an adatom diffusion from an equilibrium
site to another one on simple surfaces, such as Au(111) for
example, is straightforward since it is a straight line between
the initial and final state. In the case of more complex surfaces,
such as vicinales and reconstructed ones, the problem is more
difficult, in particular,  near steps and discommensurations,
since in these cases, the minimum energy path is no longer a
straight line between initial and final equilibrium sites. The
most reliable method is then to perform precise mappings of the
adsorption energies of the Co adatom on selected area  and then to
apply a path finding algorithm.

\begin{figure}[ht]
  \begin{center}
  \includegraphics[width=16cm]{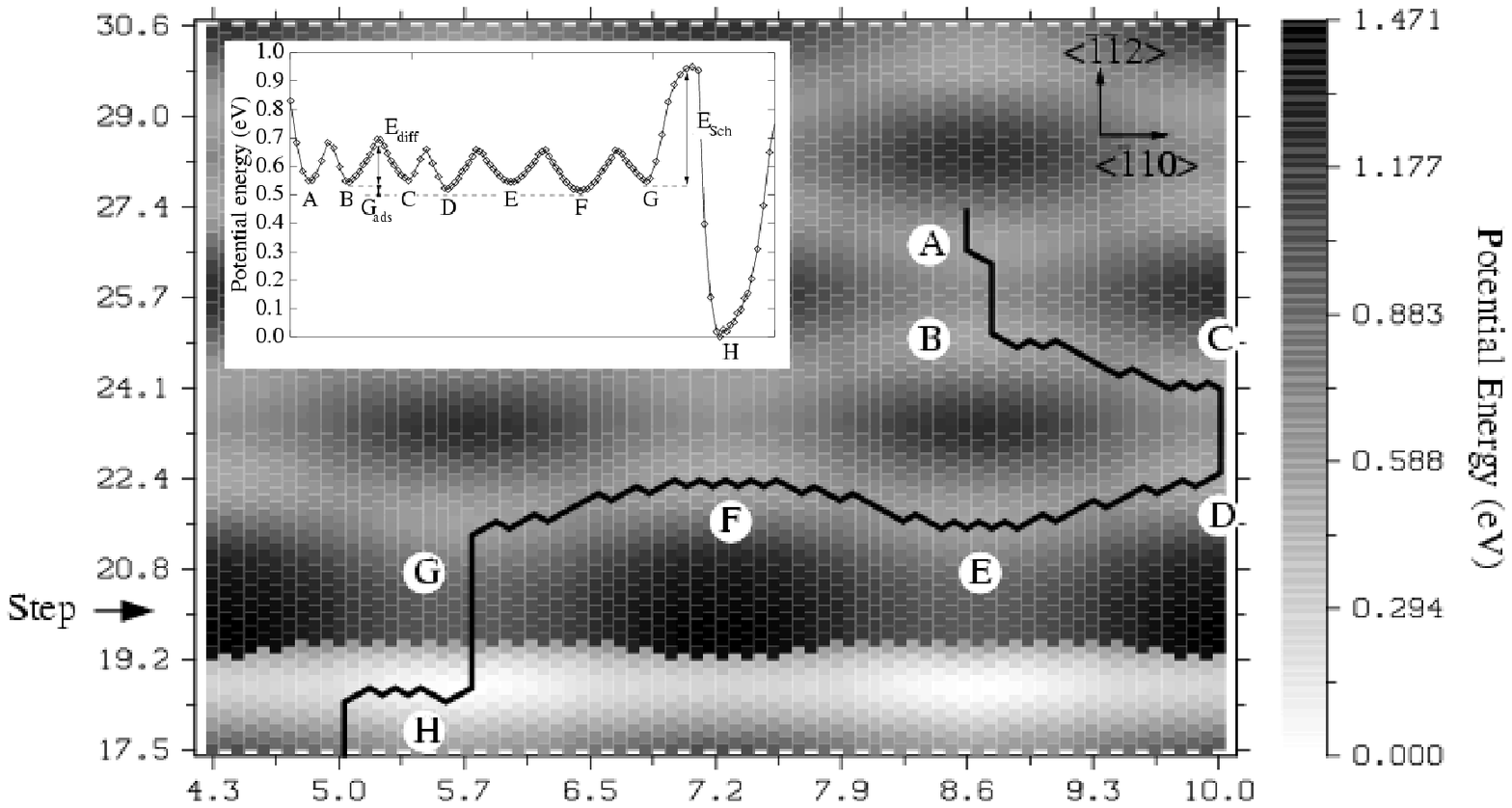}
  \caption{\label{fig:map}Energy map of a Co adsorption. The black line is the minimum
      energy path from A to H.}
  \end{center}
\end{figure}

Figure \ref{fig:map} display the energy map of a Co adsorption in
the step area. For each point, only the Co (x,y)-coordinates are
strained at fixed values during the quenching procedure. The step
of the mappings has been set to 0.100 \AA\ along the $<1\bar{1}0>$
direction and 0.087 \AA\ along the $<\bar{1}\bar{1}2>$ direction.
The Dijkstra's path finding algorithm has been applied in order to
determine the minimum energy path \cite{dijkstra1959}.

Searching for simplicity, among numerous atomistic events that
have been tested by the Molecular Dynamic, very few of them have
been selected for the KMC simulations. The dependence of $\Delta
E_i$ is represented by $\Delta E_i=E_{diff}+ n . E_{Co-Co} +
\delta_{loc} $ where $n$ is the number of Co first neighbors and
$\delta_{loc}$ is zero over the all surface but for some specific
sites where the Co energy landscape is locally modified because of
the heterogenities of the substrate. The numerical values of
parameters are given in table \ref{KMCparameters}. The quantity
$\delta_{loc}$ is describes the energies in specific sites and
will be given further as $\Delta G_{ads}$, $\Delta G_{ex}$,
$\Delta E_{ex}$. The number of atomic events and consequently the
number of parameters are reduced to a minimum in order to stress
the driving mechanisms. The surface diffusion barrier reported in
fig. \ref{fig:map} ($E_{diff}=0.15$ eV) is higher than the one
reported in Ref. \cite{goyhenex2001} ($E_{diff}=0.12$ eV ). The
origin of such a discrepancy comes from the fact that in Ref.
\cite{goyhenex2001}, the calculations have been performed on a
$22\times \sqrt{3}$ reconstructed surface ; in this case  the
stress relief induced by the reconstruction lowers the diffusion
energy of adatoms \cite{brune1995,ratsch1997,schroeder1997}. We
take $E_{diff}=0.12$ eV, which should be more correct for our
problem.

\begin{table}[htbp]
  \begin{center}
      \begin{tabular}{l|c}
        \hline\hline
        Parameter   & Value  \\ \hline \hline
        $\nu_0$     & $3.28\times10^{12}$ Hz\\
        $E_{diff}$  & 0.12 eV\\
        $E_{Co-Co}$ & 0.44 eV \\ \hline\hline
      \end{tabular}
      \caption{\label{KMCparameters} KMC parameters obtained by
      mean of Molecular Dynamic calculations. The attempt rate $\nu_0$ is calculated from
      ref. \cite{bulou2003} and the diffusion energy $E_{diff}$ can be found in
      \cite{goyhenex2001,liu2002}.}
  \end{center}
\end{table}

For the KMC simulations, the surface is $500\times500$ site wide
with periodic boundary conditions. The deposition is treated as a
random event which has a probability that is adjusted with respect
to the deposition rate as suggested in \cite{jensen1994}.

A KMC simulation with no specific sites, i.e., of a deposition on
a virtual homogeneous surface with the parameters of table
\ref{KMCparameters} is done to compare our KMC model with the RE
model for homogeneous nucleation \cite{venables1972,brune1998a}.
The variation of the critical cluster density with the temperature
is shown in fig.\ref{HOM_ADS_INS}c. A good agreement has been
found for the two growth regimes:

\begin{figure}[ht]
  \begin{center}
    \includegraphics[width=16cm]{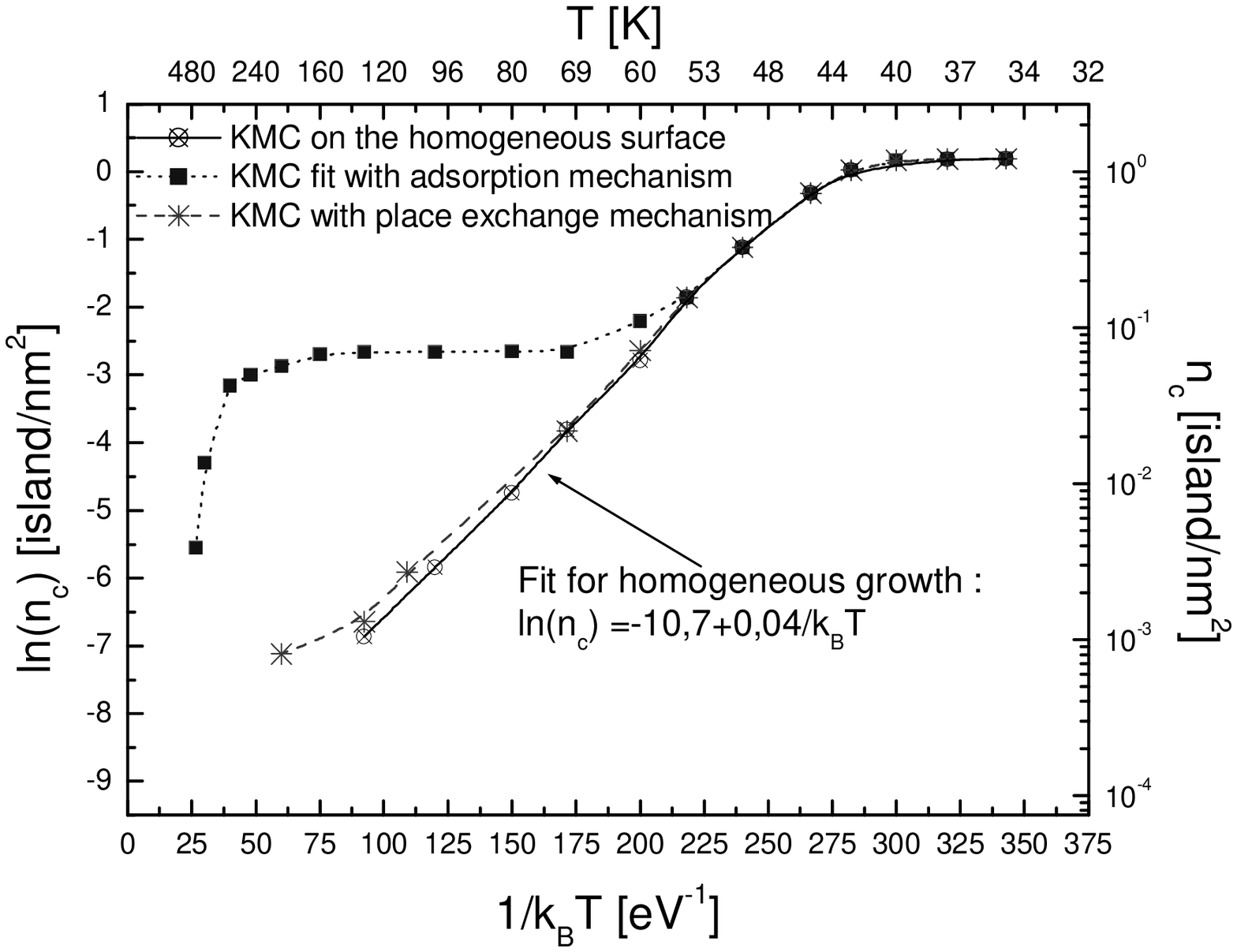}
    \caption{\label{HOM_ADS_INS} KMC result for the growth with different
models: ($\otimes$) : homogeneous surface ; ($\blacksquare$) :
surface with favored site with the favored adsorption mechanism ;
($\ast$) : surface with favored site with favored place exchange
mechanism.}
  \end{center}
\end{figure}

First, for the low substrate temperatures, the cluster density
does not depend on the temperature: this is the "post-nucleation"
regime \cite{brune1998a} (critical cluster size $i^*=0$) where the
adatoms diffusion is negligible compared to the deposition rate.

Second, for high temperatures the critical cluster density is
given by eq.\ref{i=1}. This is a regime with a critical cluster
size $i^*=1$, where the diffusion of adatoms is significant and
where dimers are stable. Using eq.\ref{i=1}, a fit of the
numerical data from KMC enables us to deduce the diffusion energy.
We find $120 \pm 2$ meV while the input $E_{diff}$ was $120$ meV
which reveals a good agreement.

For simplicity, the influence of the Schwoebel barriers yielded by
the surface steps on the adatoms diffusion is neglected in our
model. This approximation is vindicated by the fact that the
presence of the favored adsorption sites lessens the mean free
path of the adatoms which is actually lower than the steps width
as long as we have an ordered growth. Indeed, the steps have
roughly no influence over the organization while the adatoms mean
free path is smaller than the steps width. For higher temperature,
i.e. for temperature higher than $T_e$, the mean free path of
adatoms is longer than the steps width. However, as this appends
for temperatures above than $T_e$ (about 300 K), the steps are not
of a great influence on the diffusion as the energy of the
Schwoebel barrier (0.4 eV) is much lower than the energy of the
adatoms to leave the favored sites (about 0.8 eV). The temperature
threshold above which the steps are expected to play a role has
been estimated by testing a KMC simulation with a virtual surface
whose the steps involve an infinite Schwoebel barrier. It was
found that the temperature threshold is higher than $T_e$ which
indicates that our approximation is valid in order to give a good
description of the ordered growth regime.

The effect of the discommensuration lines is simplified by
introducing a set of favored sites on the KMC model surface. Each
of those sites is located near the step edge (the possibility of a
repulsive diffusion barrier to cross the discommensuration line as
suggested in \cite{goyhenex2001} is neglected). According to the
atomically resolved STM images \cite{repain2002a} of the structure
of Au(788) which show inserted Co atoms in the gold surface layer,
two favored sites are separated by 7 atomic sites in a dense
direction and the density of the favored sites is 1/200.

\begin{figure}[ht]
  \begin{center}
    \includegraphics[width=16cm]{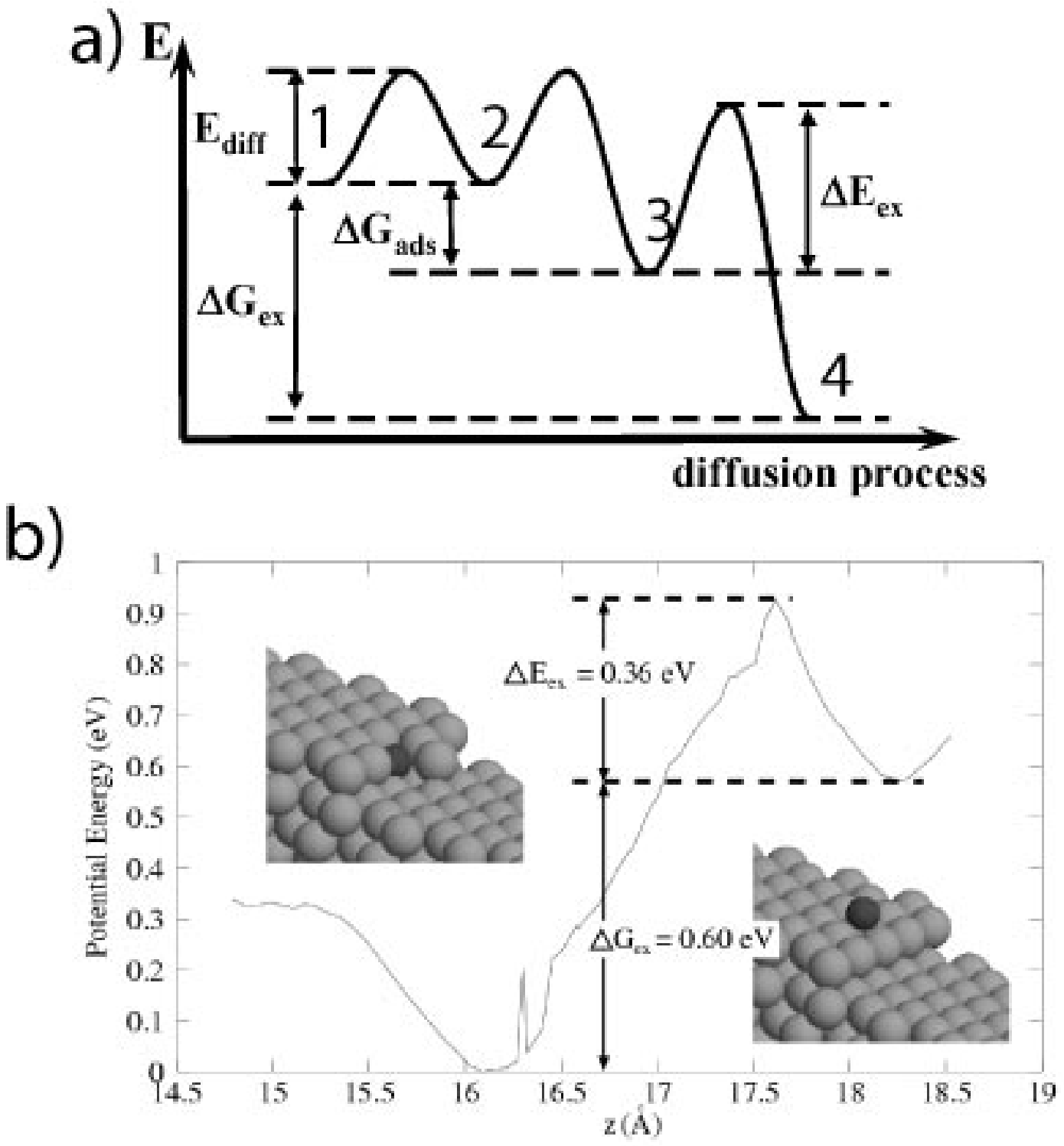}
    \caption{\label{fig:exchange} a) Energetic models used in
    the KMC simulations for the mechanisms in the favored sites:
    the adatom diffuses from normal sites (1 and 2) to the favored
    site (3 and 4). It has first an adsorption gain $\Delta
    G_{ads}$ (3) with respect to the normal sites then it can
    exchange with a gold atom (4) jumping over the exchange
    barrier $\Delta E_{ex}$. The energy gain compared to a normal site
    is $\Delta G_{ex}$ b) QMD calculation for the place
    exchange mechanism at the
    edge of a step.}
  \end{center}
\end{figure}

In the favored sites, two different mechanisms can be considered:
first a favored adsorption, which was foreseen by MD calculation
on the discommensuration line on a Au(111) surface in
\cite{goyhenex2001} and second an exchange of the adsorbed $Co$
atom in place of the underlying Au atom, which was observed by
atomically resolved STM \cite{repain2002a}. The favored adsorption
mechanism involves only one additional parameter to the
homogeneous surface parameters: the energy gain of the adatom in
the site $\Delta G_{ads}$ measured with respect to a normal
adsorption site (see position 3 in fig.\ref{fig:exchange}a). The
exchange mechanism is described with two additional parameters:
the activation barrier for the exchange $\Delta E_{ex}$ and  the
energy gain of the atom $\Delta G_{ex}$ when it is inserted in the
gold surface layer (see position 4 in fig.\ref{fig:exchange}a).
The three additional parameters have been estimated by MD
calculations. Figure \ref{fig:exchange}b displays the  minimum
energy path for a Co-Au exchange near the step. Only the Co
z-coordinate has been strained during the quenching procedure.

\subsubsection{Choice of the mechanisms for the ordered growth}

In order to prove the need for the two mechanisms described above,
we test successively the KMC simulations with only one of the two
mechanisms. A first try with the favored adsorption mechanism
leads to $\Delta G_{ads}=0.7$ eV in order to fit the thermal
variation of the maximum cluster density. A qualitative agreement
between the experiment and the KMC simulations is found. This
simulation is very close to the RE calculation described above as
they both correspond to the same mechanism. The best fit is obtain
with the same energy barrier to leave the favored site
($G_{ads}^{KMC}+E_{diff}=E_t=0.82$ eV) in both KMC and RE
calculation. It is worth noticing that the fit for low temperature
is better with KMC than with the RE model. The temperature $T_o$
is particularly well reproduced with this model ($T_o^{KMC} \simeq
75$ K, $T_o^{STM} \simeq 70$ K). The end of the plateau is much
better reproduced by KMC than by RE calculation. This is due to
the influence of the dimer bonds: in RE calculation the
possibility for a dimer to break is neglected although the dimer
binding energy ($0.52$ eV) is small with respect to the
temperatures in the end of the plateau (around $T = 300$ K).

However, the value of $\Delta G_{ads}=0.7$ eV cannot be explained
by a favored adsorption mechanism. Indeed, this value is in strong
contradiction with the MD calculations, which indicates a value of
$0.1$ eV \cite{goyhenex2001} on a flat reconstructed surface.
Although the presence of the step is not included in the MD
calculation of \cite{goyhenex2001}, the correction is not expected
to be so large. Such a value should correspond to a more complex
mechanism, such as a place exchange.

We test this hypothesis with a KMC simulation with the favored
place-exchange alone. The energy barrier to reach the favored site
is fixed to $\Delta E_{ex} = 0.32$ eV so that the place exchange
does not occur before 200 K according to the VT-STM experiment.
The gain of the Co atom when it is inserted is $\Delta G_{ex} =
0.5$ eV and thus the energy to leave the favored sites is $0.82$
eV. In the limits of the explored temperature range, we have only
seen few deviation of the growth regime from the growth on an
homogeneous growth. The small deviation at about 200 K is
explained by the activation of the favored place exchange
mechanism. The qualitative disagreement can be explained as
follows: when the adatoms have sufficient energy to diffuse on a
distance equal to the mean distance between favored sites, their
thermal energy is still too low to enable them to reach the
favored site, due to the access barrier $\Delta E_{ex}$ (we can
remark that this value is very low in comparison to typical values
for the place exchange on a flat, unreconstructed surface). As a
consequence, the density keeps on decreasing as the favored sites
cannot play any role. When the thermal energy of the adatoms
overpass the place exchange barrier then the nucleation becomes
heterogeneous but there is still less than one dot per favored
site. We can conclude that the favored place exchange mechanism
cannot explain alone the experimental results. This is consistent
with our low coverage images for various temperatures which do not
show the place exchange mechanism below $200$ K.

\subsection{Interplay between adsorption and place exchange mechanisms.}

\subsubsection{Final model}

\begin{figure}[ht]
  \begin{center}
    \includegraphics[width=16cm]{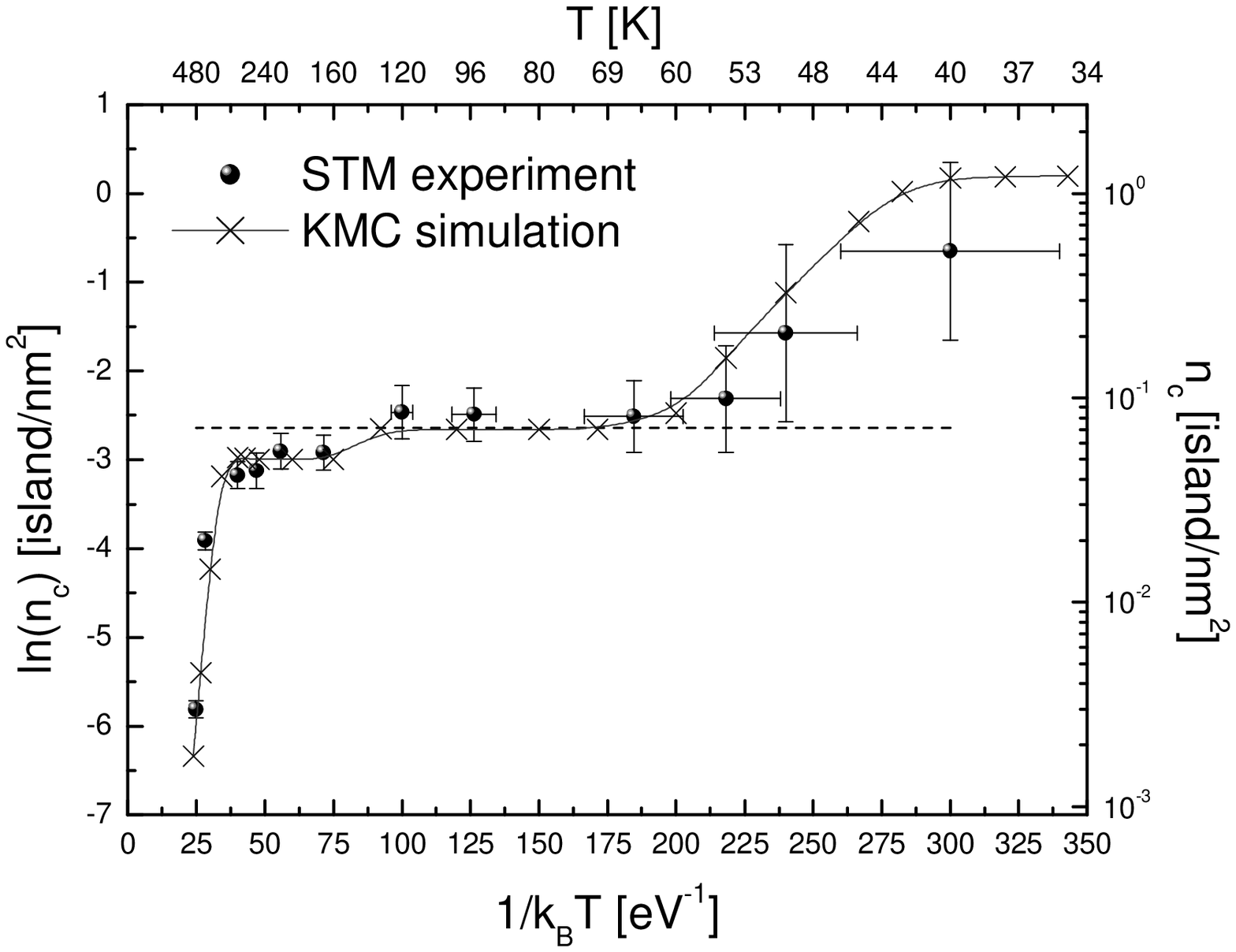}
    \caption{\label{kmc_glob} KMC simulation of the growth of Co
      nanodots on Au(788). A good fit of the experimental data for
      the maximum cluster density evolution with temperature is
      obtained with a model combining the favored adsorption and
      favored place exchange mechanisms in the favored sites.}
  \end{center}
\end{figure}

A realistic fit of the experimental data is achieved with the
combination of the two mechanisms inside the favored sites. The
thermal variation of the critical cluster density
fig.\ref{kmc_glob} is found with the parameters that are
summarized in the table \ref{fitparameters}. This table gives also
a set of parameters calculated by QMD. These parameters are
calculated on various surfaces as we do not know precisely the
atomic structure of the Au(788) surface. This allows us to make a
comparison with the parameters found in KMC calculations. The
adsorption gain $\Delta G_{ads}$ was found to be 0.28 eV in KMC.
This high value with respect to the KMC must be due to the
influence of the step edge. Indeed, on the discommensuration line
on the Au(111) surface, every sites along the line is equivalent.
This is not the case on the Au(788) surface which indicates that
the step edge changes the adsorption energies and make a more
favorable site. The values found by the QMD for the exchange
mechanism are of the same order as the values of the KMC although
the discommensuration lines is not included in this calculation.
The fact that the exchange takes place at a precise is due to
kinetic effect. Indeed the exchange and the favored adsorption
take place at the same place. The adsorption gain in this site
increases the residence time of the adatoms on this sites which
increases the exchange probability in the same way.

\begin{table}[ht]
    \begin{center}
      \begin{tabular}{l|ccc}
        \hline\hline
        Parameter            & KMC     & \multicolumn{2}{c}{Molecular Dynamic}\\
                             & &Au(788)          &Au(111)\\
                             & &non reconstructed&reconstructed\\ \hline \hline
        $E_{diff}$           & 0.12 eV & 0.15 eV & 0.12 eV \\
        $\Delta G_{ads}$     & 0.28 eV &         & 0.10 eV\\
        $\Delta G_{ex}$      & 0.78 eV & 0.60 eV & \\
        $\Delta E_{ex}$      & 0.32 eV & 0.36 eV & \\
        \hline\hline
      \end{tabular}
      \caption{\label{fitparameters} KMC parameters for the favored
       adsorption and the exchange mechanisms to fit the experimental data.
       For comparison various MD parameters are given. Calculation were done on a Au(788) non reconstructed surface
       and on a Au(111) reconstructed surface (see \cite{goyhenex2001,goyhenex2002}).
       The exchange on the Au(788) surface was performed at 1 rows of the step edge.
       The exchange and adsorption on Au(111) were performed on the top of the discommensuration line.}
    \end{center}
\end{table}

\begin{figure}[ht]
  \begin{center}
    \includegraphics[width=16cm]{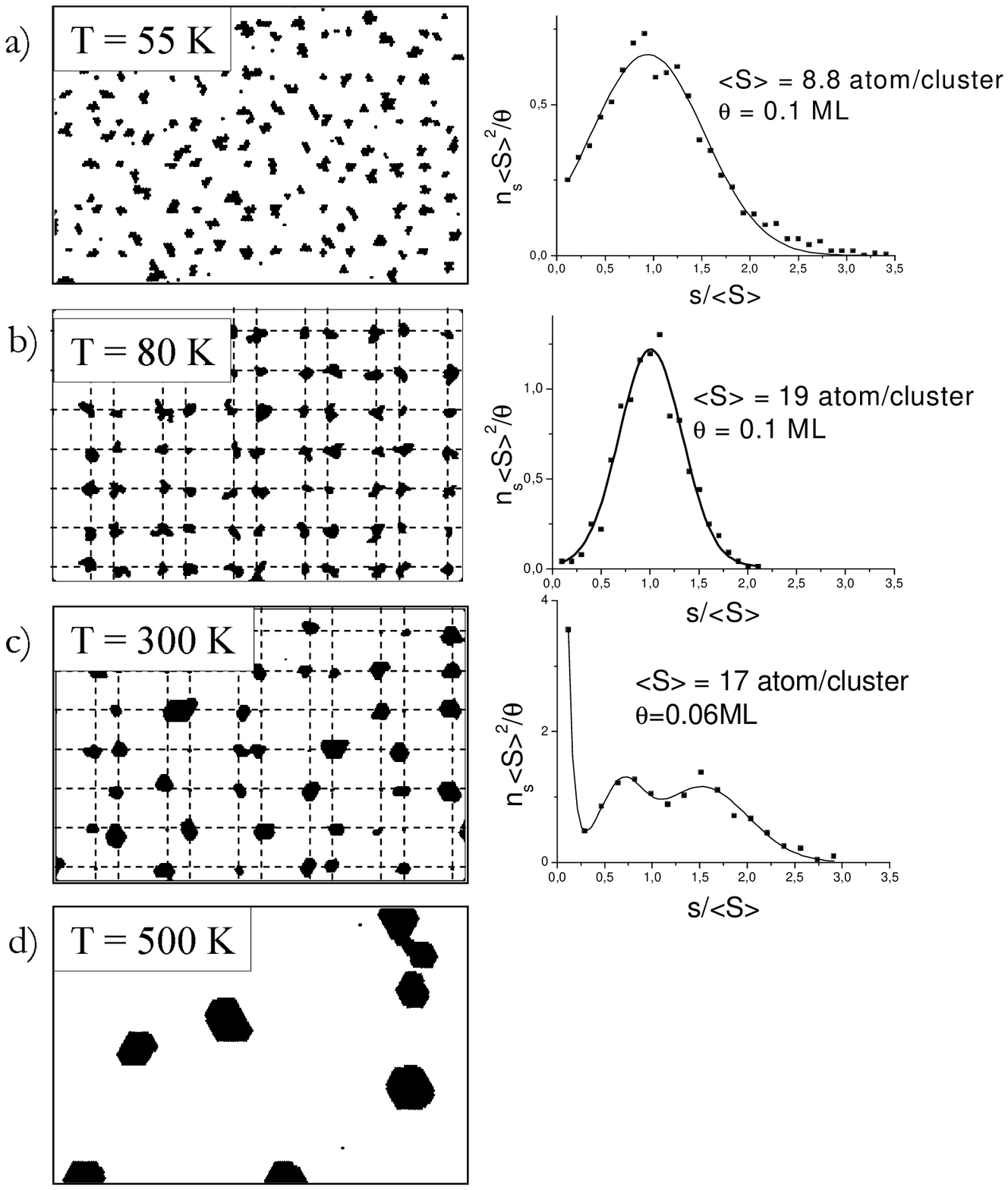}
    \caption{\label{kmcfig} KMC images and size distributions of the
      simulation of the growth of
      Co nanodots on Au(788). They show the good agreement
      with STM images. Images are $20\times40$ nm (a-c) or
      $50\times 75$ nm (d) wide and are taken for
       a) $T=55$ K, b) $T=80$ K, c) $T=300$ K and d) $T=450$ K. In the
      image e), the asymetric shape of the clusters that we have found
      with STM is not reproduced due to the roughness of our KMC model
      which does not take into account the difference between the diffusion
      along A and B steps.}
  \end{center}
\end{figure}

A plateau with exactly one dot per favored site is found in the
temperature range which lies from $75$ K to $150$ K. This plateau
can only be explained with the favored adsorption mechanism as no
place exchange occurs for so low temperatures. The KMC images
shown in figure \ref{kmcfig} are in good agreement with STM
experiments and confirm that the best condition for a long range
order with a narrow cluster size distribution is a sample
temperature around 130 - 150 K.

Above $150$ K, the energy gain is no more sufficient to stabilize
the adatoms and the clusters density decreases. Above $200$ K, the
place exchange becomes efficient in the simulations, so the
cluster density stops to decrease and a rough organization is
maintained up to $300$ K, i.e., the ratio $n_c$ of the number of
cluster per favored sites is such as $1<n_c<2$. The most important
point is that in this temperature range, the order is still
present on the surface (see fig. \ref{kmcfig}c) but the number of
defect (empty favored site, coalesced neighbor dots...) is more
important than in the previous regime. This fact is particularly
evident for the simulation at $T = 300$ K, i.e. the bimodal size
distribution is pretty well reproduced. The origin of this size
distribution is the weakness of the bonds between the Co atoms. As
a consequence, the critical size (biggest unstable cluster) of a
cluster on a favored site is 2. This explains why the growth of
the dots in the favored sites is strongly inhomogeneous: some dots
grow faster an lead to big clusters when they coaless with the
neighboring cluster.

One may wonder why in the $200$ K - $300$ K temperature range
there is an organization whereas in the simulation including only
the exchange mechanism with the same parameters we do not find any
organization in the same temperature range. This can be explained
by the fact that the evolution of the cluster density is not
simply explained by the sum of two different mechanisms but a
combination. Above $200$ K, even if the favored sites cannot
stabilize adatoms with just the favored adsorption mechanism, the
residence time of the adatoms is increased in these sites, with
respect to the normal sites. As a consequence, the exchange
probability is increased in the same way.

\subsubsection{Discussion}

This interplay between the two mechanisms is a key which explains
the large temperature range for the organized growth. Indeed place
exchange mechanism is not favorable for the organized growth due
to the large place exchange barrier but provide a strong energy
gain in the favored site. On the opposite, the favored adsorption
mechanism lead to the organization but the low energy gain cannot
maintain an organization in a large temperature range. The good
combination of these two mechanisms provides the organization for
a large temperature range. We have determined the four pertinent
parameters $E_{diff}$, $\Delta G_{ads}$, $\Delta E_{ex}$ and
$\Delta G_{ex}$ in our system. In order to generalize this model
to other systems, we now discuss which relation are required for
these parameters in order to obtain an ordered growth for a large
temperature range.

First of all, the width of the ordered growth regime is determined
by $E_{diff}$ for the temperature $T_o$ and by $E_t = \Delta
E_{ex}+\Delta G_{ex}-\Delta G_{ads}$, the energy barrier for the
removal process of the place exchange (when a Au adatom exchanges
with a Co inserted atom), for the temperature $T_e$. A key to
obtain a regular organization on the whole temperature range is
the interplay between the two mechanisms. As a consequence, the
removal process for the adsorption mechanism should not occur
before the place exchange is possible. From an energetic point of
view this means that the gain of the favored adsorption mechanism
$\Delta G_{ads}$ must be higher or of the same order than the
place exchange barrier $\Delta E_{ins}$. However the absolute
value of these energies is not a key as the temperature at which
the interplay occurs is not very important. In the case of Co on
Au(788) $\Delta G_{ads}$ is just of the order of $\Delta E_{ins}$.
That is why we only have a rough organization in the 170 K - 300 K
temperature range. As the interplay between the two mechanisms is
not perfect, we can see a small "accident" in the evolution of the
critical clusters density with temperature (see fig.
\ref{kmc_glob}).

\section{Conclusion}

The ordered growth of Co nanodots on the Au(788) surface is proved
to be due to the presence of favored site where two microscopic
mechanisms occur: a favored adsorption and a place exchange
mechanism. The combination of these two mechanisms leads to the
organization of the Co nanodots for a temperature that ranges from
$60$ K to $300$ K. A rectangular array of mono-disperse cobalt
nanodots is then yielded with a great regularity. These
experimental results open the way to study the physical properties
of nano-sized ferromagnetic particules with integrating technics
like Magneto-Optical Kerr Effect. With such a mono-disperse array
of nanodots integrating technics can be used to deduce the
physical properties of a single particle.

The numerical simulations, in spite of many approximations,
provide us an accurate insight about the fundamental mechanisms of
the growth of Co/Au(788). A mechanism at the origin of the
organized growth is a place exchange. For that kind of mechanism,
the removal process is very hard as it corresponds to an atomic
rearrangement in the surface layer. It was found to be helped by a
favored adsorption mechanism. Indeed, due to the energy gain of
the adatoms in these sites, the place-exchange probability is
increased because the mean residence time of atoms is large. This
explains why the growth of Co on Au(788) is organized with a so
large temperature range. Further experiments and calculations
about the deposition of other materials on Au(788) should be of
interest to confirm our nucleation and growth scenario.

We gratefully acknowledge C. Priester and G. Prevot for the
enlightening discussions.

\end{document}